\documentclass[twocolumn,showpacs,preprintnumbers,amsmath,amssymb,prl]{revtex4}

\usepackage{graphicx}
\usepackage{dcolumn}
\usepackage{bm}

\begin{document}

\preprint{APS/123-QED}

\title{Origin of the $\sim$150 K Anomaly in LaOFeAs; Competing Antiferromagnetic
Superexchange Interactions, Frustration, and Structural Phase Transition
}


\author{T. Yildirim$^{1,2}$}\email{taner@nist.gov}%
\affiliation{%
$^{1}$NIST Center for Neutron Research, National Institute of Standards and
Technology, Gaithersburg, Maryland 20899, USA
\\$^{2}$Department of Materials Science and Engineering, University of
Pennsylvania, Philadelphia, PA 19104, USA}%

\date{\today}

\begin{abstract} 

From first principles calculations we find that the nearest and  next
nearest neighbor superexchange interactions between  Fe ions  in LaOFeAs 
are large,  antiferromagnetic (AF), 
 and
give rise to a frustrated magnetic ground state which consists of  two 
interpenerating  AF square sublattices with M(Fe)=0.48$\mu_B$. 
The system lowers its energy further by removing 
the frustration via  a structural distortion. 
These results successfully explain the magnetic and structural phase
transitions in LaOFeAs recently observed by neutron scattering.  
 The presence of competing strong antiferromagnetic exchange 
interactions and the frustrated ground state suggest that  magnetism and superconductivity in 
doped LaOFeAs may be strongly coupled, much like in the high-T$_c$ cuprates.

\end{abstract}

\maketitle

\begin{figure}
\includegraphics[height=5cm]{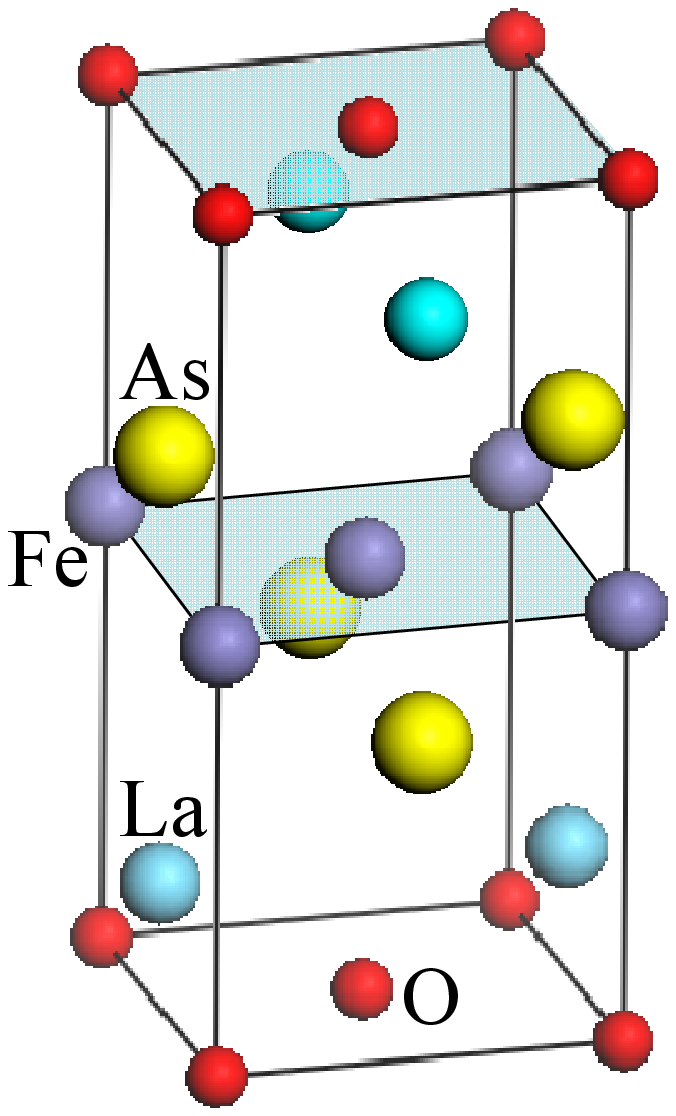} 
$\;\;$
\includegraphics[height=5cm]{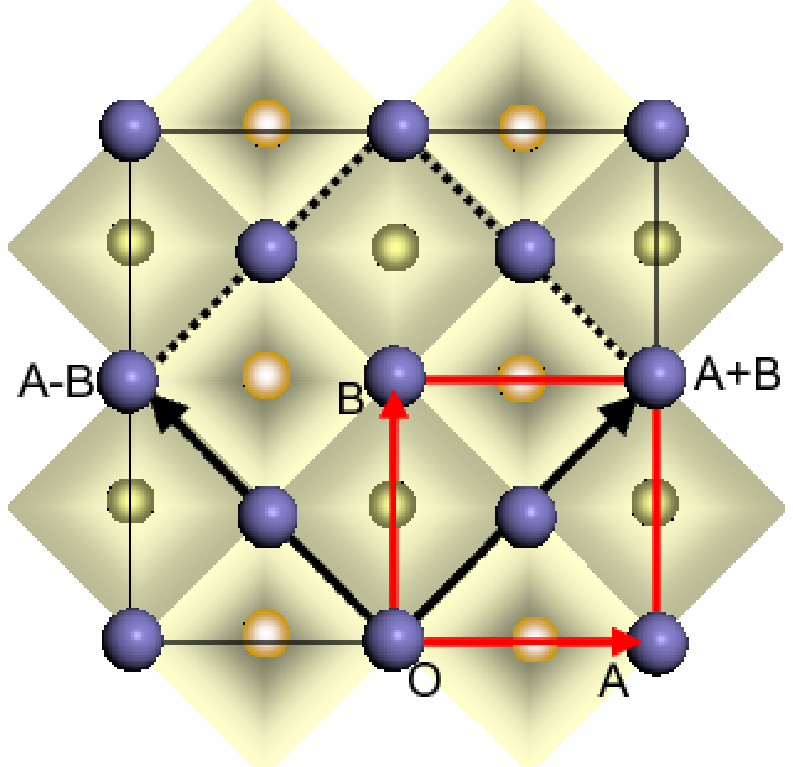} 
\caption{
(color online)
(a) The crystal structure of LaOFeAs in space group P4/nmmm with origin choice 1. 
(b) Top view of the FeAs-plane and the relations between primitive and 
$\sqrt{2}\times\sqrt{2}$ supercell used in our calculations. The dark and light shaded areas
indicate the As atoms below and above the Fe-square lattice, respectively.
}
\label{fig:figure1}
\end{figure}

The recent discovery of superconductivity at T$_c$'s up to 50~K in layered rare-earth  (R) transition 
metal (Tm) 
pnictide(Pn)-oxide  quaternary compounds ROTmPn 
(R=La, Ce, Sm, Tm=Mn, Fe,Co, Ni, Pn=P, As)\cite{kamihara,sm_43k,ce_41k,pr_52k}
has sparked enormous interest in this class of materials. These are the first non-copper based
materials that exhibit superconductivity at relatively high temperatures upon electron 
(O$_{1-x}$F$_x$)\cite{kamihara,sm_43k,ce_41k,pr_52k} and hole doping
(La$_{1-x}$Sr$_x$)\cite{sr_doped} of their nonsuperconducting parent compounds, just like 
high-T$_c$ cuprates.
Clearly, the understanding of electronic, magnetic, and structural properties of the parent compound
LaOFeAs is the key to determining the underlying mechanism that makes these materials superconduct upon 
electron/hole doping. 

Recent theoretical studies suggest conflicting results varying from 
LaOFeAs being a nonmagnetic metal near a ferromagnetic or antiferromagnetic instability\cite{singh,xu,haule} to a 
simple antiferromagnetic semimetal\cite{cao,ma}.
From optical measurements and density functional calculations, 
it was also suggested that LaOFeAs has
an antiferromagnetic spin-density-wave (SDW) instability due 
to Fermi-surface nesting\cite{dong}. 
Experimental studies including resistivity and magnetic susceptibility  
show a small but very clear
anomaly near 150~K in LaOFeAs\cite{kamihara,dong}. 
The origin of this anomaly has been very recently determined by neutron
scattering studies\cite{cruz,mcguire}. It has been found that LaOFeAs exhibits an
antiferromagnetic long-range  ordering with a small 0.35$\mu_{B}$ per Fe moment  followed 
by a small structural distortion\cite{cruz}. However there
is no proposed microscopic theory that explains the origin of the observed stripe-like
AF spin decoration and the  structural distortion. 
It is also not clear if the magnetic and structural phase transitions
 are related to each other.  Finally, given the fact that both the cuprates and LaOFeAs  exhibit antiferromagnetic
ordering, one wonders  how strong and what kind of magnetic spin-fluctuations are present in the 
2D Fe-square lattice of
LaOFeAs. 

In this letter, from accurate all-electron  density functional calculations we try to answer some of these
questions. We find that both  nearest neighbor  (nn) and  next
nearest neighbor (nnn) superexchange interactions between Fe ions  in the square lattice are very large,
comparable to each other and more importantly they are 
antiferromagnetic. This forces the Fe spins along the square-diagonal to order antiparallel,
resulting two interpenetrating square AF sublattices. Since in this configuration we have
 one parallel and one antiparallel alignment of the spins
along the square axes, the nearest neighbor superexchange interaction is totally frustrated.
The system lowers its energy  by removing this frustration via a structural distortion that
makes the two sides of the square-lattice inequivalent. These results including the 
magnetic moment of the Fe ions and the degree of structural distortion, are in excellent
agreement with the recent neutron data\cite{cruz}.  Therefore,  we have a nice working
microscopic theory that explains the details of both the magnetic and structural properties of the
undoped parent compound LaOFeAs. Our theory  also
brings our attention to the presence of the strong competing antiferromagnetic interactions in this
class of materials. Even though electron doping seems to destroy the long-range magnetic order, the
short range spin fluctuations will be always present and probably play an important role in the
superconducting phase, much like the high T$_c$ cuprates. 
In fact, a theory has been already proposed for the superconductivity 
mediated by antiferromagnetic spin fluctuations
in LaOFeAs\cite{mazin}.

\begin{figure}
\includegraphics[width=3.5cm]{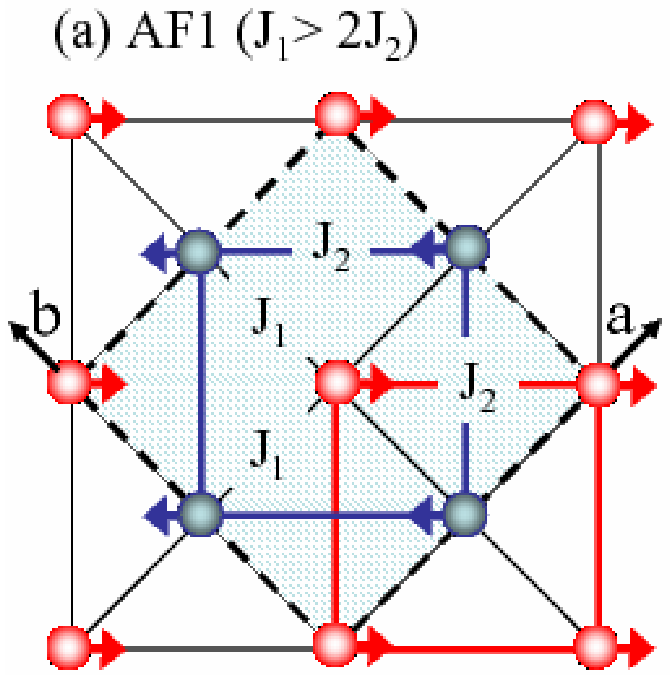}
$\;\;$
\includegraphics[width=3.5cm]{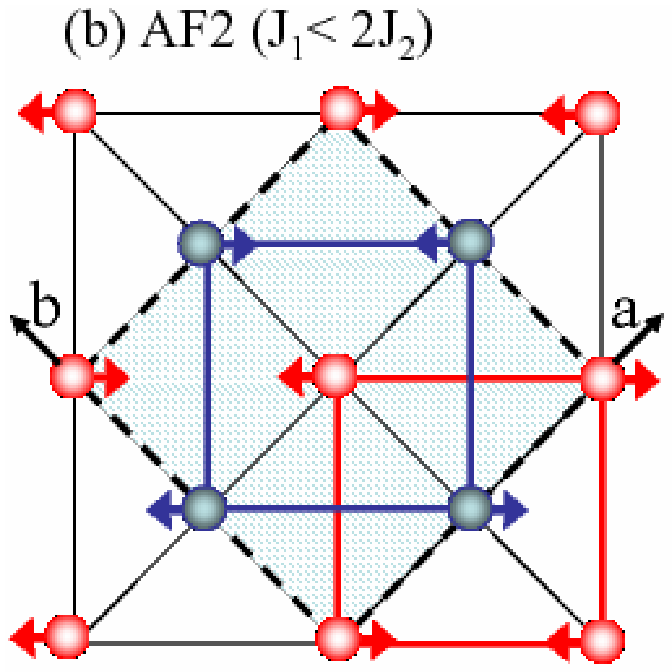}
\caption{
(color online)
Two antiferromagnetic configurations considered in this study. Left panel shows the AF1 
configuration where nearest neighbor spins are always aligned antiparallel. 
 Right panel shows the AF2 configuration where the next nearest
neighbor spins (i.e.,J$_2$) are always aligned antiparallel.  
 }
\label{fig:figure2}
\end{figure}

The calculations were done using the full-potential linearized augmented planewave (FP-LAPW) method,
 one of the most accurate methods available in electronic structure
 calculations\cite{wien2k,exciting}. We also used the ultrasoft pseudo potential planewave (PW)
 method\cite{pwscf} for cross checking  of our results and for phonon calculations. 
  We considered 
 $\sqrt{2}\times\sqrt{2}$ supercell of the primitive cell of LaOFeAs which is shown in
 Fig.~1. In order to determine the true ground state, we have considered four different
 cases. These are non-magnetic (NM, i.e., no spin polarization), ferromagnetic (F)  and the 
 two different antiferromagnetic spin
 configurations shown in Fig.~2. The first one of the antiferromagnetic configurations is AF1 
 where the nearest neighbor spins are  antiparallel to each other. 
  The second antiferromagnetic configuraiton, AF2, is shown in Fig. 2b.
 In AF2 the Fe spins along the square diagonal are aligned antiferromagnetically. The AF2 spin
 configuration can be considered as two interpenerating simple square AF sublattices (red and blue
 sublattices in Fig.2b). We note that since each Fe ion is at the middle of a square AF lattice, the
 mean field at each spin site is zero. Hence one sublattice can be rotated
 freely with respect to the other sublattice without costing any energy. For this reason the AF2
 spin-configuration is fully frustrated. This can be also seen by the fact that for a given square, we
 have always one parallel (J$_1$S$^2$)  and one antiparallel (-J$_1$S$^2$) aligned spin pair,
 cancelling each others contribution to the total energy. In frustrated magnetic systems, it is known that the
 frustration is almost always removed by either a structural distortion or by thermal and quantum
 fluctuations\cite{yildirim_bct,shlee}. From the classical energies of AF1 and AF2, one sees that the
 AF2 spin configuration is stabilized when  $J_2 > J_1/2$.

\begin{figure}
\includegraphics[width=4cm]{En_vs_M.eps}
$\;\;$
\includegraphics[width=4cm]{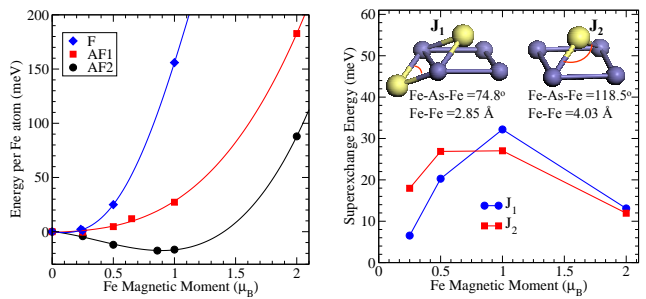}
\caption{
(color online)
(a) The total energy per Fe atom versus magnetic moment for F, AF1 and AF2 spin-configurations,
indicating AF2 is the only ground state of the system. (b) The superexchange parameters for nn and nnn
Fe ions obtained from the energies of F, AF, and AF2 configurations.  
 }
\label{fig:figure3}
\end{figure}

In order to determine which spin configurations among NM, F, AF1, and AF2, is the ground state, 
we have carried out FP-LAWP total energy calculations for each case. 
Since in spin-polarized calculations  it is
very easy to get a local minimum, we followed a different strategy. In our calculations we fixed the
magnetic moment per Fe ion and then  scan the total energy as a function of Fe-magnetic moment. 
Our results are summarized in Fig. 3. The zero of energy is taken as the
M=0 case (i.e. NM calculation). From Fig. 3, it is clear that LaOFeAs has only one magnetic ground
state which is AF2. The Ferro spin-configuration always results the highest energy regardless  the
Fe-magnetic moment. Similarly AF1 ordering always yields energies higher than the NM case.
For the AF2 ordering, we see that
the energy minimum occurs near the fixed moment calculation with M=1. Repeating calculations where
magnetization is not fixed, we obtained the optimum magnetic moment  as M=0.87 $\mu_B$ per Fe. 
As we discuss below in detail, the Fe magnetic moment is further reduced almost by a half when the structure is
allowed to distort to remove the magnetic frustration. We note that the magnetic moment obtained from FP-LAPW method 
is significantly smaller than those obtained from pseudo-potential  PW based calculations which give moments around $M$=2 to 2.5$\mu_B$.
We also note that there is no indication of any ferromagnetic interactions  present in LaOFeAs which is clear from
Fig.~3. In fact, the system prefers to reduce its magnetization to zero if spins are forced to be aligned
ferromagnetically. Therefore,  we conclude that the main superexchange interactions in LaOFeAs system is antiferromagnetic. 
The fact that AF2 has lower energy than AF1  indicates that the next nearest neighbor interaction is also
antiferromagnetic and  satisfies the phase boundary $J_2 > J_1/2$.

\begin{table*}
\caption{The symmetries and energies (in meV) of the optical phonons of LaOFeAS in P4/nmm and P2/c phases.
The  energies of the IR-active modes are  taken from Ref.\onlinecite{dong}. 
The $^{*}$ indicates a significant disagreement! The animations of these modes can be found at 
http://www.ncnr.nist.gov/staff/taner/laofeas}
\begin{center}
\begin{tabular}{|ccc|ccc|ccc|ccc|ccc|} \hline \hline
\multicolumn{15}{|c|}{
$\Gamma (P4/nmm )$ = 2 A$_{1g}$ (IR) + 4 A$_{2u}$(IR) + 4 E$_{u}$ (IR) + 4 E$_{g}$(R)  + 2 B$_{1g}$ (R) } \\ 
\multicolumn{15}{|c|}{
$\Gamma (P2/c )$ $\;\;\;\;\;\;\;$= 2 A$_{g}$ (IR) + 4 A$_{u}$(IR) + 8 B$_{g}$ (R) + 8 B$_{u}$ (IR)} \\ \hline
P4/nmm & P2/c & IR  & P4/nmm & P2/c & IR &  P4/nmm & P2/c & IR &  P4/nmm & P2/c & IR & P4/nmm & P2/c & IR \\ \hline
E$_{u}$ 7.3 & 7.4-7.5&  -- & 
A$_{2u}$ 9.9&  10.1& 12.1 & 
E$_{g}$ 14.0& 14.1-14.2& -- & 
E$_{g}$ 17.6& 17.7-17.8& -- & 
A$_{1g}$ 22.1& 22.3& -- \\
A$_{1g}$ 24.9 & 25.1& -- & 
B$_{1g}$ 26.6 & 26.9& -- &  
A$_{2u}$ 31.2 & 31.6& 30.9 & 
E$_{u}$ 33.7 & 34.0-34.1 & 33.2 & 
B$_{1g}$ 35.2& 35.6& -- \\
E$_{g}$ 35.6& 35.9- 36.1& -- & 
 E$_{u}$ {\bf 34.3}& {\bf 34.6- 34.7}& {\bf 42.0$^{*}$ } & 
A$_{2u}$ 49.1& 49.1& 53.8 &  
E$_{g}$ 51.6& 51.8-52.6& -- &  & & \\ \hline \hline
\end{tabular}
\end{center}
\end{table*}

In order to gain a better insight into the nature of the superexchange interactions present in Fe-square
lattice of the LaOFeAs system, we map the calculated total energies of the  F, AF1 and AF2
configurations shown in Fig.3a to a simple Heisenberg like model $H= \sum_{i,j} J_{i,j} M_{i} M_{j} $
for a given fixed Fe moment $M_{i}$. For fully localized spin-systems this is a perfect thing to do but
for the case of LaOFeAs this is only an approximation. Nevertheless, the calculated $J$s should be a good
indication of the superexchagne interactions present in the system. The Fig. 3b shows the $J_1$ and $J_2$
obtained from the energies of the F, AF1 and AF2 at given magnetic moment. It is clear that both $J_1$
and $J_2$ are quite large and positive (i.e. antiferromagnetic). $J_2$ is always larger than $J_1/2$J
and therefore AF2 structure is the only ground state for any given moment of the Fe ion. By looking at the
superexchange paths for J$_1$ and J$_2$ (shown in insets to Fig.3), we notice that  the Fe-As-Fe angle
is around $75^{\rm o}$ and $120^{\rm o}$ for nn and nnn Fe-pairs, respectively. Hence it makes sense that
the 2nd nn exchange interaction is as strong as the nn exchange because the angle is closer to the optimum
value of $180^{\rm o}$.  
It is quite surprising and also very interesting that there are strong and competing antiferromagnetic superexchagne
interactions in LaOFeAs system that  result in  a totally frustrated AF2 spin configuration. This is very
similar to the magnetic ground state of the cuprates where the AF ordered 2D square lattices of the adjacent planes are
frustrated\cite{yildirim_bct}.

\begin{figure}
\includegraphics[width=8cm]{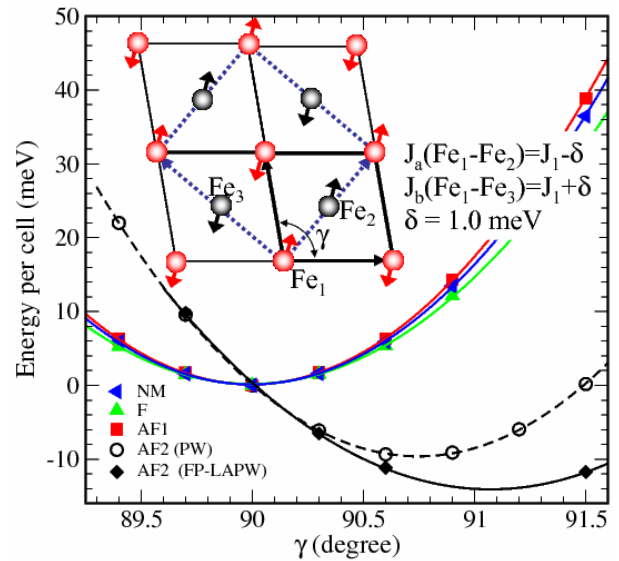}
\caption{
(color online)
The total energy per cell versus the angle $\gamma$ for non-magnetic (NM), Ferromagnetic (F) and two
antiferromagnetic (AF1 and AF2) spin-configurations. Note that only the AF2 spin configuration
yields structural distortion. The inset shows that as $\gamma$  increases,  the ferromagnetic aligned
Fe ions (i.e., Fe$_1$-Fe$_2$) get closer while the antiferomagnetically aligned ions (i.e., Fe$_1$-Fe$_3$) move
apart, breaking the four-fold symmetry and thus the degeneracy of the d$_{xz}$ and d$_{yz}$ orbitals.
 }
\label{fig:figure4}
\end{figure}

We next discuss the implication of the magnetically frustrated AF2  configuration on the
structural distortion recently observed by neutron scattering\cite{cruz}. It is a common observation that when the
system is magnetically frustrated, the frustration is usually lifted by a structural distortion\cite{shlee}. In the
case of LaOFeAs, the frustration is due to the parallel and antiparallel alignment of the spins along
the sides of a square. Hence one expects to see a structural distortion which brings the two Fe spins closer
to each other while for the other side, they move apart from each other. In fact this is exactly what
Cruz {\it et al.} observed in their neutron scattering experiments. In order to demonstrate the removal of
frustration by structural distortion, we calculated the total energy of the AF2 spin configuration as a
function of the $\gamma$ angle as shown in the inset to Fig.~4. When $\gamma=90^{\rm o}$, we have the original tetragonal cell. Once the
$\gamma $ deviates from $90^{\rm o}$, the original $\sqrt{2}\times\sqrt{2}$ structure (shown as dashed
line) is no longer tetragonal but orthorhombic (i.e., the cell length along a and b axes are no longer
equal). 
The total energy versus $\gamma$ plot shown in Fig.4 clearly indicates that the structure is indeed
distorted with  $\gamma = 91.0^{\rm o}$, which is in good agreement with the experimental value of $90.3^{\rm o}$. 
We note that  PW calculations (dotted line) are in excellent agreement with the FP-LAPW method except that the
magnetization of the Fe ion comes out very large. From the FP-LAPW method we get M=0.48 $\mu_B$ which is in
excellent agreement with the experimental value of $0.35 $ $\mu_B$.  We also considered two types of
AF2 where the spins along the short axis are aligned parallel or antiparallel. These two configuration
are no longer equivalent. From the energies of these configurations we deduce the new
superexchange interactions along the a and b directions which are given in Fig.~4. The net energy gain byg
the structural distortion is about 12 meV per cell, which is of the same order as the temperature at
which this phase transition occurs. 

In order to make sure that the structural distortion is driven mainly by frustration removal and not by
other effects, we have also calculated total energy versus $\gamma$-angle for other spin configurations
including the non-magnetic case. We note that for $\gamma=90^{\rm o}$, the orbitals $d_{xz}$ and
$d_{yz}$ are degenerate and therefore one may think that the system is subject to  symmetry lowering for reasons
similar to those in a Jahn-Teller distortion. However as shown in Fig. 4, we do not see any distortion for any of 
the NM, F, and AF configurations. Therefore, the  experimentally observed structural distortion is due
to removal of the frustration generated by the AF2 ordering.

\begin{figure}[b]
\includegraphics[width=6cm]{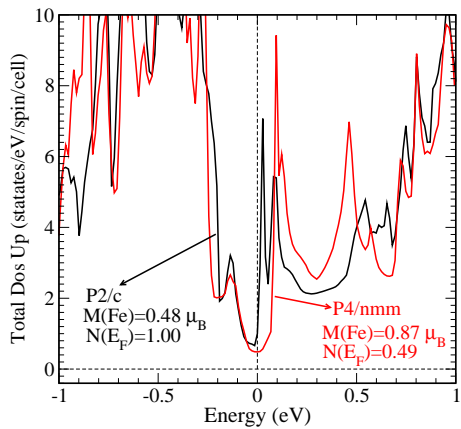}
\caption{
(color online)
The total electronic DOS of LaOFeAs  for high (P4/nmm) and low (P2/c) symmetry phases.  
 }
\label{fig:figure5}
\end{figure}

We now  briefly discuss the effect of the structural distortion on the electronic
structure and the zone center phonons. The details will be published elsewhere. Fig. 5 shows the electronic density of states
before and after the structural distortion with the AF2 ordering in $\sqrt{2}\times\sqrt{2}$ structure. 
The N(E$_F$) in both phases are quite small compared to previous calculations in which incorrect AF ordering was considered. 
It is apparent that the distortion has significant effect both on the N(E$_F$) and the Fe magnetization.
Interestingly, the N(E$_F$) is doubled while the magnetization is reduced by half due to structural distortion.
The increase in the N(E$_F$) is consistent with the resisitivity measurement which shows that LaOFeAs exhibits metallic 
behavior after the transition\cite{dong}. Finally, we note the existing several sharp  Van Hove like kinks in the DOS. In particular the
distortion brings one of these kinks just next to the Fermi energy. Hence, with a small electron doping, it is quite possible to increase the 
N(E$_F$) significantly. The effect of such electron doping is under study and will be 
reported elsewhere.

Finally we explain why Dong {\it et al.} did not see any evidence of the sructural phase transition 
in their optical IR measurements\cite{dong}. From the the symmetry decomposition of the optical phonons in both P4/nmm and
P2/c phases (see Table~1), we note that the distortion does not introduce any new
IR active modes but rather just splits the doubly-degenerate modes into non-degenerate ones.
However the splitting is quite small; the largest is around 0.2 meV. This explains why no new modes appear in the
optical measurements after the transition. We also note that the agreement for the 
energies of the zone center phonons with IR data is not as good as one expects. In particular, the $E_{u}$ mode observed at 
42 meV is calculated to be 35 meV, a significantly lower value. Interestingly, this particular mode has a strong
temperature dependence\cite{dong}.  We checked that the disagreement is not due to anharmonic phonons. 
The calculated phonons are harmonic unlike
those observed in the MgB$_2$ superconductor\cite{taner_mgb2}.We hope that our observation will
motivate others  to look at the zone -center phonons carefully to understand the discrepency.

In conclusion, we have presented a first-principles study of
 the superexchange interactions between Fe ions and 
their effect on the magnetic  and structural
properties  of the parent compound LaOFeO of the newly discovered high temperature 
superconductor LaO$_{1-x}$F$_{x}$FeAs. 
 The competing strong antiferromagnetic exchange interactions and the frustrated
 ground state suggest that LaOFeAs has many common magnetic properties with the undoped parent component
 of the high-T$_c$ cuprates.

The author acknowledges useful discussions with 
C. Cruz, R. L. Cappelletti, Q. Huang, J. W. Lynn and W. Ratcliff.

\end{document}